\documentclass{iau}
\usepackage{graphicx}
\title[Proceedings IAU Symposium No. 294, 2013] 
{Geometrical information on the solar shape:
 \\ high precision results with SDO/HMI}
\author[Xiaofan Wang \& Costantino Sigismondi]   
{Xiaofan Wang$^1$ \and Costantino Sigismondi$^2$}
\affiliation{$^1$Key Laboratory of Solar Activity, National Astronomical Observatories, \\
Chinese Academy of Sciences, email: {\tt wxf@nao.cas.cn} \\[\affilskip]
$^2$ICRA/Sapienza Universit{\`a} di Roma, Ateneo Pontificio Regina Apostolorum,
\\email: {\tt sigismondi@icra.it}}
\pubyear{2013}
\volume{294}  
\jname{Solar and Astrophysical Dynamos and Magnetic Activity}
\editors{A.G. Kosovichev, E.M. de Gouveia Dal Pino \& Y. Yan, eds.}
\begin{document}
\maketitle
\begin{abstract}
The uncertainty of measurement of solar diameter is depending on the observational time scale.
Full-disc images of SDO/HMI and the images from ground observations in Huairou Solar
Observing Station have been analyzed to get the values of solar diameter.
The satellite observations reach a very high
precision, but the absolute image scale still need to be calibrated.
The solar oblateness is a more challenging
measurement than the diameter, since the signal amplitude is a few
milli-arcseconds. It is a relative measurement, then not affected
by the pixel scale calibration required by the diameter measurement. But the results
are strongly dependent on the state of instrument such as focus plane deformation
and on the calculation process.

\end{abstract}
\section{Introduction}
The ground based solar diameter measurement have been performed since
very long time (e.g. Thuillier, G. et al., 2005; Lefebvre, S. et al., 2006; Sigismondi, 2011).
Generally speaking, the study of solar diameter should be based on long term series of
observations since it is a quite stable quantity, representative of the whole star,
or of its outer shells.
A global observation network to monitor the solar diameter can be realized by different cross-calibrated
full-disc solar telescopes.
When we compare the solar diameter observation results
acquired by Huairou Solar Observing Station from 2006 to 2009 (not shown in this paper) with the results acquired by
SDO/HMI in one year (2011), we find the solar diameter results based on ground observation
largely affected by Earth atmospheric conditions.
Conversely the space result accuracy is strongly dependent on the orbital parameters (Fig. 1).

\begin{figure} \begin{center}
 \includegraphics[width=5.1in]{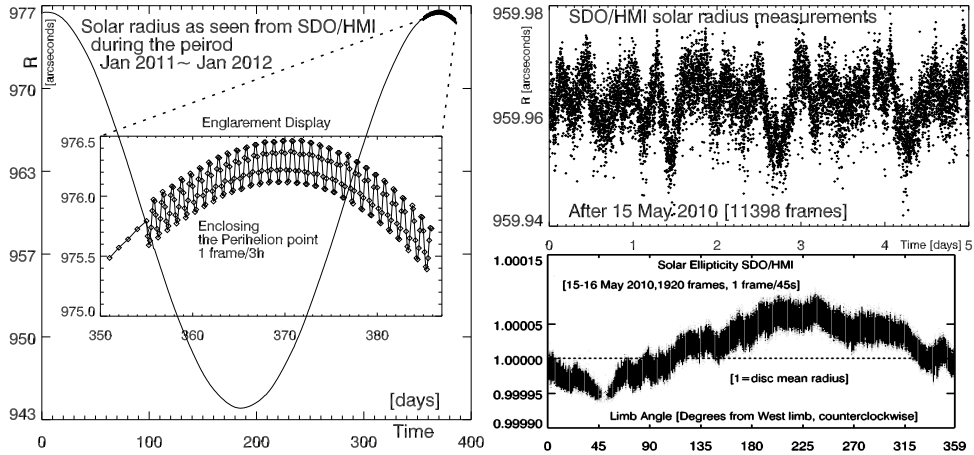}
 \caption{The apparent solar radius affected by the Earth orbital motion and the satellite orbital motion (left image with enlargement).
 The solar radius appears as a fluctuating signal when
 the orbital motions are removed (upper-right image), and there is still some orbital motion
 residuals.
 The apparent shape of solar disc calculated from 1920 full-disc
 images on very quiet activity period (15-16 May 2010) shows slightly departures from the sphere
 (lower-right image), but some invisible solar structures or instrumental problems may influence
 the results seriously.}
   \label{fig1}
\end{center}
\end{figure}

\section{Measurements of the solar diameter}

The solar diameter (radius) observations on the space are continuous
when the instrument is in normal operation. We used the HMI continuum images in the following
discussion since its focal plane devices reach very high performance.
We choosed 365 images from one year data set.
But we used higher cadence (1 frame/3hour) around the perihelion,
in order to check the effects of SDO Earth-rotation synchronous orbit to the apparent solar radius measures.
Further, we normalize the apparent solar radii
to the distance of 1 AU according to the HMI images informations (the distance satellite-Sun seems to be very accurately determined),
and to its one-year average. In this way we can only see variations below one year of timescale down to 45 s.
This value varies smoothly (at the moment we are not sure it is due to the approximations used in orbital calculation
 or it is due to a real measurement of orbit).
But the pixel scale and the focus step
parameters (see e.g. Khun et al., 1998; Emilio et al., 2012)
are observational conditions which also are depending on the satellite orbit,
and their precision seem to be worse than the nominal ones.
Solar oblateness (ellipticity) is a difficult topic, more related to the dynamics and physics of the solar interior.
The absolute pixel calibration, orbit information,
and focus step are no longer essential, but flat field,
focus plane deformation, and CCD small tip-tilt can affect seriously this measure.
Rolling the telescope is a fair solution for such instrumental
problems (Emilio et al., 2007 and 2012). To locate weighted inflexion points of the solar limb we used, as Emilio et al. (2012), polar coordinates transform of limb stripes with a bilinear interpolation, while
Kuhn et al. (1998) decomposed the solar limb position and brightness into
a linear combination of Legendre polynomials.

\section{Conclusions}

\noindent{\bf(a)} The solar radius averaged over one year in our analysis is 959.963 $\pm$ 0.005 arcsec
(1 $\sigma$). It is 0.333 arcsec larger than the standard value (959.63 arcsec) and 2 $\sigma$ smaller than 960.12 $\pm$ 0.12 arcsec of Emilio et al. (2012).
Higher accuracy could
depend on better orbit information,
better data set (single wavelength),
and absolute pixel scale calibration using Mercury and Venus transits (Sigismondi and Wang, 2013);
Averages over shorter timescales can be done.
{\bf (b)} The oblateness is just a preliminary result
(roughly estimate, 0.048 $\pm$ 0.03 arcsec (1$\sigma$))
since the deformation of focus plane is unknown.

\end{document}